\documentclass[10pt]{article}
\pdfoutput=1
\usepackage{graphicx,amssymb,amsfonts,amsmath,amssymb,amscd,amstext, mathrsfs,color,xcolor}
\makeatletter

\@addtoreset{equation}{section}
\newcommand{\be}{\begin{eqnarray}}
\newcommand{\ee}{\end{eqnarray}}
\newcommand{\ba}{\begin{array}}
\newcommand{\ea}{\end{array}}
\newcommand{\nn}{\nonumber}

\makeatletter \@addtoreset{equation}{section} \makeatother

\begin{document}
\vspace{1cm}
\begin{center}
~\\~\\~\\
{\bf  \LARGE Entanglement Entropy of Compactified Branes  and Phase Transition}
\vspace{1cm}

                      Wung-Hong Huang\\
                       Department of Physics\\
                       National Cheng Kung University\\
                       Tainan, Taiwan\\

\end{center}
\vspace{1cm}
\begin{center}{\bf  \Large ABSTRACT}\end{center}
We first calculate  the holographic entanglement entropy of M5 branes on a circle and see that it  has phase transition during decreasing the compactified radius.  In particular, it is shown that the 
entanglement entropy scales as $N^3$.  Next, we investigate the holographic entanglement entropy of D0+D4 system on a circle and see that it scales as  $N^2$ at low energy, likes as a  gauge theory with instantons.   However, at high energy it transforms to a phase which scales as $N^3$, like as  M5 branes system.  We also present the general form of holographic entanglement entropy of   Dp, ${\rm D_p+D_{p+4}}$ and M-branes on a circle and see some simple relations between them.  Finally, we present an analytic method to prove that they all have phase transition from connected to disconnected surface during increasing the line segment of length $\ell$ which dividing the space. 
\vspace{2cm}
\begin{flushleft}
*E-mail:  whhwung@mail.ncku.edu.tw\\
\end{flushleft}
%%%%%%%%%%%%%%%%%%%%%%%%%%%%%%%
\newpage
%%%%%%%%%%%%%%%%%%%%%%%
\tableofcontents
%%%%%%%%%%%%%%%%%%%%%%%
%%%%%%%%%%%%%%%%%%%%%%%
\section{Introduction}
The entanglement entropy $S_A$  in quantum field theories or quantum many body systems is a non-local quantity. It is defined as the von Neumann entropy $S_A$ of the uced density matrix when we trace out (or smear out) the degrees of freedom inside a d-dimensional space-like submanifold B in a given (d+1)-dimensional QFT, which is a complement of A.  Physically, the observer who is only accessible to the subsystem A will feel as if the total system is described by the uced density matrix $\rho_A$
\be 
{\rm \rho_A=Tr_B~ [\rho_{total}]}
\ee
The entanglement entropy of the subsystem A is defined by 
\be 
{\rm S_A=-Tr_A~ [\rho_A~log\, \rho_A]}
\ee
which measures how the subsystems A and B are correlated with each other.   This origin of entropy looks analogous to the black hole entropy. {  Indeed, this was the historical motivation of considering the entanglement entropy in quantum field theories \cite {Hooft,Bombelli, Snicki,Cardy1,Cardy2, Cardy3}.   Entanglement entropies have played an important role in condensed matter physics  \cite {Amico, Eisert, Kitaev}, quantum gravity \cite{Raamsdonk,Raamsdonk1005, Myers1304, Balasubramanian}, and quantum information \cite{Horodecki}. }

As the d + 1 dimensional conformal field theories have $AdS_{d+2}$ gravity duals  \cite {Maldacena, Gubser,Witten}, S. Ryu and T. Takayanagi \cite {Ryu,Ryu0605,Takayanagi} conjectu that the entanglement entropy between the regions A and B is proportional to the classical area of this surface, 
\be
S_A = {1\over 4G_N^{(d+2)}}\int_\gamma d^d\sigma \sqrt{G_{ind}^{d}}
\ee
where $G_N^{d+2}$ is the d+2 dimensional Newton constant and $G_{ind}^{d}$  is the induced string frame metric on $\gamma$.  {  Some literatures  attempted to  prove above formula \cite{Fursaev,Lewkowycz,Casini}.  Recent review of calculating entanglement entropy and holographic entanglement entropy can be found in \cite{Nishioka}. }

In non-conformal theories the dilaton are in general not constant. A natural generalization to the corresponding ten dimensional geometries was proposed by 
\be
S_A = {1\over 4G_N^{(10)}}\int_\gamma d^8\sigma~e^{-2\Phi}\sqrt{G_{ind}^{8}}
\ee
It is natural to generalize above relation to the eleven dimensional M-brane geometries 
\be
S_A = {1\over 4G_N^{(11)}}\int_\gamma d^9\sigma \sqrt{G_{ind}^{9}}
\ee
{ Holographic entanglement entropy had been extensively used to investigate the phase transition in superconductor and other matters, for example the early papers in \cite{Albash,Cai}. The confinement and deconfinement transition can also be studied by using the holographic entanglement entropy \cite{Klebanov,Hartnoll}.} 

Klebanov et al. conside the holographic entanglement entropy of ``slab"  (the subspace defined by $-\ell/2 < x < \ell/2$, where $x$ is one of the spacial coordinates) of  Dp branes compactified on a circle \cite{Klebanov,Pakman,Ben-Ami}.   They had found that there are two types of minimal surfaces: connected and disconnected ones and the D4 and D3 systems have a phase transition, confinement deconfinement transition, between these two types of solutions. For small $\ell$ the connected solution dominates the computation of entanglement entropy, while for large $\ell$  the disconnected solution becomes prefer.  { On the other hand, Hartnoll et al.\cite{Hartnoll} conside the  holographic entanglement entropy for  the  hypersurface $\Sigma$ consisting of two parallel infinite spatial hyperplanes separated by a distance $L$.  In the confined phase the entaglement entropy is found to be $S=c\,Vol(\Sigma)+div$.  On the other hand, in the  deconfined phases $S=c\,L\,Vol(\Sigma)+div$.

While both approaches are very general and could be applied to any system, in this paper we only  follow Klebanov method \cite{Klebanov} to investigate the systems of  M brane and Dp branes which are compactified on a circle and to find the phase transition therein.

Note that there is another quantity, called as Renyi entropy \cite{Renyi,Headrick1006, Klebanov1111,Fursaev1201} which is a one-parameter generalization of entanglement entropy labeled by an integer n
\be
{\rm S^{Renyi}_A= {1\over 1-n} log\,Tr_A~ [\rho^n_A]}
\ee
and in the $n\rightarrow 1$  it uces to entanglement entropy.    The holographic method had been extended to calculate the Renyi entropy \cite{Hung1112,Belin1306} and the charged Renyi entropy that includes a chemical potential \cite{Belin1310,Belin1407}.  While the Renyi entropy can  provide us more information than entanglement entropy we only investigate  the $n\rightarrow 1$  limit in this papr.}

In section II we first investigate the holographic entanglement entropy M5 branes on a circle.  We see that it scales as $N^3$  \cite{Gubser9602, Klebanov9604}, and has the phase transition  during decreasing the compactified radius.  Next, we  investigate the holographic entanglement entropy of D0+D4 system on a circle and see that it scales as  $N^2$ at low energy, likes as a gauge theory with instantons \cite{Douglas,Lambert}.   However, at high energy it scales as $N^3$, like as a M5 system.  In section III we present the general form of holographic entanglement entropy for Dp branes, ${\rm D_p+D_{p+4}}$ and M-branes on a circle and  see a simple relation between them.  Finally, in section IV we present an analytic method to prove that they all have phase transition from connected to disconnected surface during increasing the line segment of length $\ell$ which dividing the space. Last section is used to summarize our results and mention some future works.

After this paper is completed \cite{Quijada} appea which has  some overlaps with the material presented here. However, since we have calculated the  entropy of  D0+D4  we could study the phase transition of  M5 branes. Also we find that the entanglement entropy of compactified  M5, M2,  Dp and ${\rm D_p+D_{p+4}}$ has similar mathematic form and thus study the universal phase transition.
%%%%%%%%%%%%%%%%%%
\section {Holographic entanglement entropy of  M5 and D0+D4 on circle}
\subsection {Holographic entanglement entropy of  M5 on circle}
Near-horizon M5 matric on a circle of coordinate $w$ with radius $R_w$ is
\be ds^2= {U\over R_5} \Big[-dt^2+dx^2 +\sum_{i=1}^3 dx_i^2 \Big] + {U\over R_5}\Big(1-{b^3\over U^3}\Big)dw^2+ {R_5^2\over U^2}{dU^2\over 1-{b^3\over U^3}}+  R_5^2 d\Omega_4^2
\ee
in which $b= {4R_5^3\over 9R_w^2}$ and $R^3$ is proportional to number of M5 branes (see section 3). In terms of $U=4R_5^3/z^2$ and define  
\be z_c\equiv 3R_w\ee
 then
\be ds^2= {4R_5^2\over z^2}\Big[-dt^2+dx^2+\sum_{i=1}^3 dx_i^2+\Big(1-{{z^6\over z_c^6}}\Big)dw^2+ {dz^2\over 1-{z^6\over z_c^6}}  \Big]+ R_5^2d\Omega_4^2
\ee
On a time slice t =constant the induced metric of M5 is
\be ds_{ind}^2= {4R_5^2\over z^2}\Big[\sum_{i=1}^3 dx_i^2 +\Big(1-{{z^6\over  z_c^6}}\Big)dx_4^2+ \Big({z'^2\over 1-{z^6\over z_c^6}}+1\Big) dx^2\Big]+  R_5^2 d\Omega_4^2
\ee
Using $\int dw=2\pi R_w$, $\int dx_i =L$ and $\int \Omega_4=\omega_4$ we find that surface is 
\be 
\label{M5}
 A&=&\int_0^{2\pi R_w} dw ~\int \Pi_{i=1}^{3}dx_i\int \Omega_4~\int_{-\ell/2}^{\ell/2} dx~\sqrt {g_{ind}}\nn\\
&=&32R_5^9~2\pi R_w~L^3~2\omega_4\int_0^{\ell/2} dx~{\sqrt{z'^2+1-{z^6\over  z_c^6}}\over z^5}
\ee
{To find the profile function z(x) we can regard variable $x$ in above equation as time and the Hamiltonian becomes
\be
 H_0&=& {\delta {\cal L}\over\delta z'}z' -{\cal L}= {-1\over z^5}{1-{z^6\over  z_c^6}\over\sqrt{z'^2+1-{z^6\over  z_c^6}}}
\ee
We see that the Hamiltonian  does not depend on x and it is a constant of motion. Thus
\be
z'=\sqrt{{ (1-{z^6\over  z_c^6})^2\over H_0^2\,z^{10}}-(1-{z^6\over  z_c^6})}
\ee
Since that $z'=0$ at turning point $z=z_*$  we  have a relation
\be
 {1\over H_0^2}&=&{z_*^{10}\over 1-{z_*^6\over  z_c^6}}
\ee
Thus the profile function z(x) is described by the differential equation
\be
\label{z}
z'=\sqrt{{z_*^{10} (1-{z^6\over  z_c^6})^2\over z^{10}(1-{z_*^6\over  z_c^6})}-(1-{z^6\over  z_c^6})}
\ee}
In considering the entanglement entropy of the ``slab" in which the subspace is defined by $-\ell /2 < x < \ell/2$ we have the following relation
\be
\label{ell}
\ell/2=\int_0^{\ell/2} dx =\int_0^{z_*}dz {dx\over dz}=\int_0^{z_*}dz {z^5 \sqrt{1-{z_*^6\over  z_c^6}}\over \sqrt{{z_*^{10} (1-{z^6\over  z_c^6})^2}-z^{10}(1-{z_*^6\over  z_c^6})(1-{z^6\over  z_c^6})}}
\ee
Above relation is used to find function $ \ell (z_*)$.  Note that $z_*$  is independent of number of M5 branes $N$. 

Use the function form of $z'$ in \eqref{z}  the connected surface becomes
\be 
\label{A}
 A_C=(32R_5^9~2\pi R_w~L^3~2\omega_4)\int_0^{z_*}dz {z_*^5\sqrt{1-{z^6\over  z_c^6}}\over z^5\sqrt{{z_*^{10} (1-{z^6\over  z_c^6})}-z^{10}(1-{z_*^6\over  z_c^6})}}
\ee
To proceed we shall note that  there is disconnected surface which is defined by taking $z_*=z_c\equiv 3R_w$.  In this case we have an exact result :
\be 
\label{B}
A_D&=&(32R_5^9~2\pi R_w~L^3~2\omega_4)~ \int_\epsilon^{z_c}dz {1\over z^5}\nn\\
&=&(32R_5^9~2\pi R_w~L^3~2\omega_4)~ \Big(-{1\over 4 z_c^4}+{1\over 4 \epsilon^4}\Big)
\ee
in which $\epsilon$ is a cutoff parameter.  Let us now begin to analyze the above relation. 

1. From  \eqref{ell} we see that $z_c$ and $z_*$ are independent of $R_5$. Also, from the  surface formula \eqref{A} and  \eqref{B} we see that $R_5$ only appears as an overall factor $R_5^9$.  As $R_5^3$ is propositional to number of M5, $N$, the holographic entanglement entropy of  compactified M5 has the simple property : $S\sim N^3$  \cite{Gubser9602, Klebanov9604}.

2. From  \eqref{ell} we see that the slab length $\ell=0$ at $z_*=0$ or $z_*=z_C$.  Thus, increasing the value of $z_*$ the  slab length $\ell$ will be increasing too, which, however become zero at largest value of $z_*=z_C$ and  { there is a maximum value of $\ell$. See the figure 1 in which $0\le z_*\le 1$  and $\ell_{max}\approx 0.47$ under the scale $z_C=1$.
\\
\scalebox{0.2}{\hspace{20cm}\includegraphics{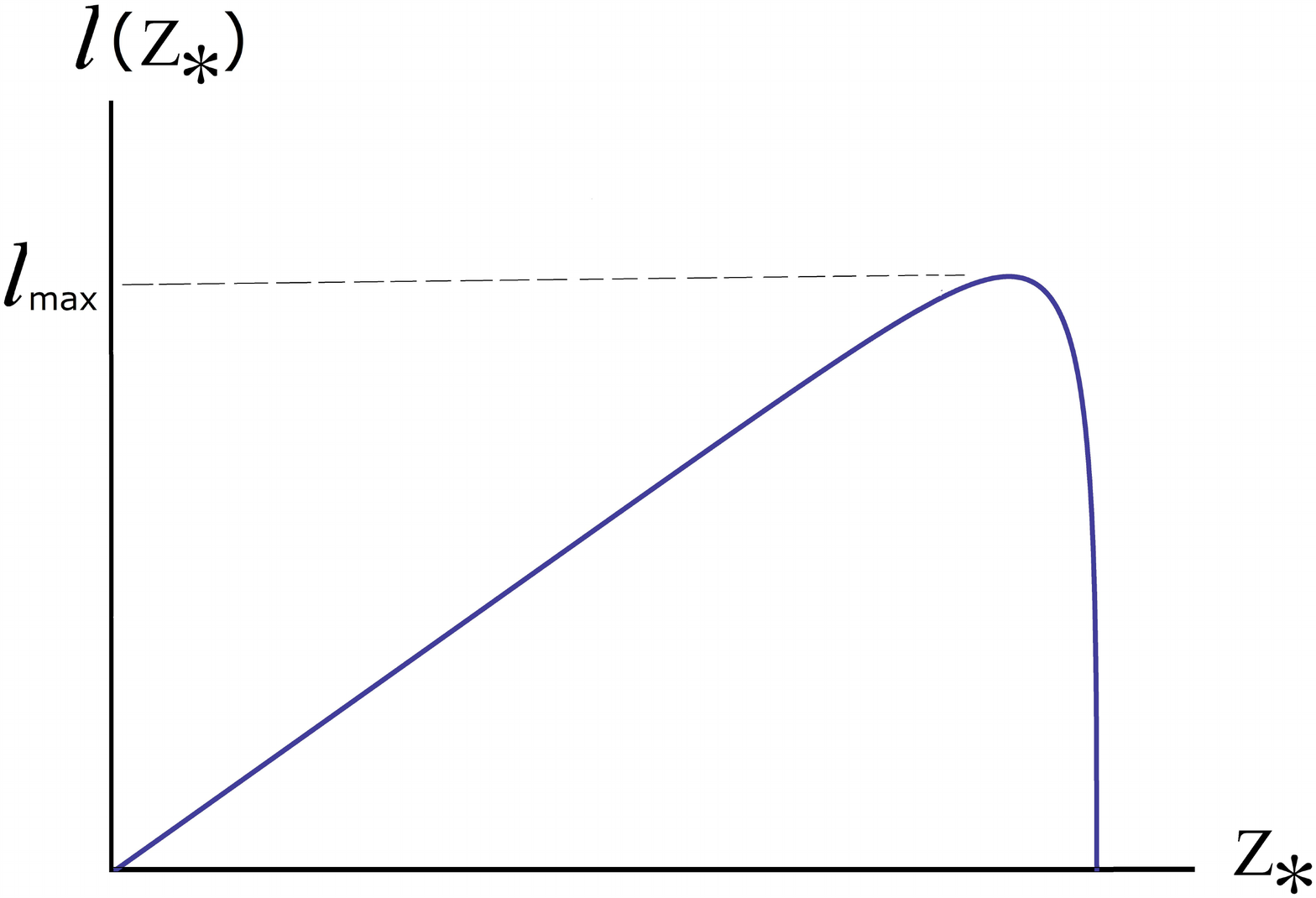}}
\\
{\it Figure 1:  Function $ \ell (z_*)$  : Slab length $\ell$ as function of  turning point $z_*$.  There is a maximum value of $\ell$}}
\\

3. From \eqref{ell}  we see that at small value of $z_* \approx 0$ and the connectioned surface becomes
\be  A_C\approx (32R_5^9~2\pi R_w~L^3~2\omega_4)~ \Big(-{1\over 4 z_*^4}+{1\over 4 \epsilon^4}\Big)
\ee
Thus $A_C-A_D<0$.  However,  increasing the value of $z_*$ the connected surface $A_C$ will be increasing too, which, will become the largest value at $z_*=z_C$. The central problem to be studied is whether before $z_*=z_C$ the connected surface could be larger then disconnected surface.  If the answer is yes then the M5 will have phase transition. 

4. In numerical analysis, we have to integrate \eqref{ell} to find the function $\ell (z_*)$ and integrate \ in {A} to find the surface function $A(z_*)$.  Then we use the two functions to obtain the surface function $A(\ell)$.   May be between some regions of  $\ell$ the disconnected surface is smaller then connected surface while between another regions of  $\ell$ the disconnected surface is larger then connected surface $A_C<A_D$.  In this case the system has phase transition at a critical value of $\ell$. This is the property that Klebanov et.el first found in the D4 and D3 systems \cite{Klebanov}. 

{ In  figure 2 we plot the connected area function  $A_C(\ell)$ and disconnected area function $A_D(\ell)$ respectively. The area functions and thus the entanglement entropy of the connected and disconnected solutions cross at at $\ell=\ell_{crit}$.  We see that $A_C\ge A_D$ when $\ell \le \ell_{critical}$. $A_C$ is maximin at $\ell=\ell_{max}$. In the intervial  $\ell_{crit}\le \ell\le \ell_{max}$ the connected area function $A_C(\ell)$ has two solution which, however, are metastable solutions \cite{Klebanov}.  And, for $\ell\ge\ell_{crit}$ the phase is governed by the disconnected solutions. Since the value of $A_C(\ell_{crit})$ and $A_D(\ell_{crit})$ are different the transition is first order.
\\
\\
\scalebox{0.25}{\hspace{18cm}\includegraphics{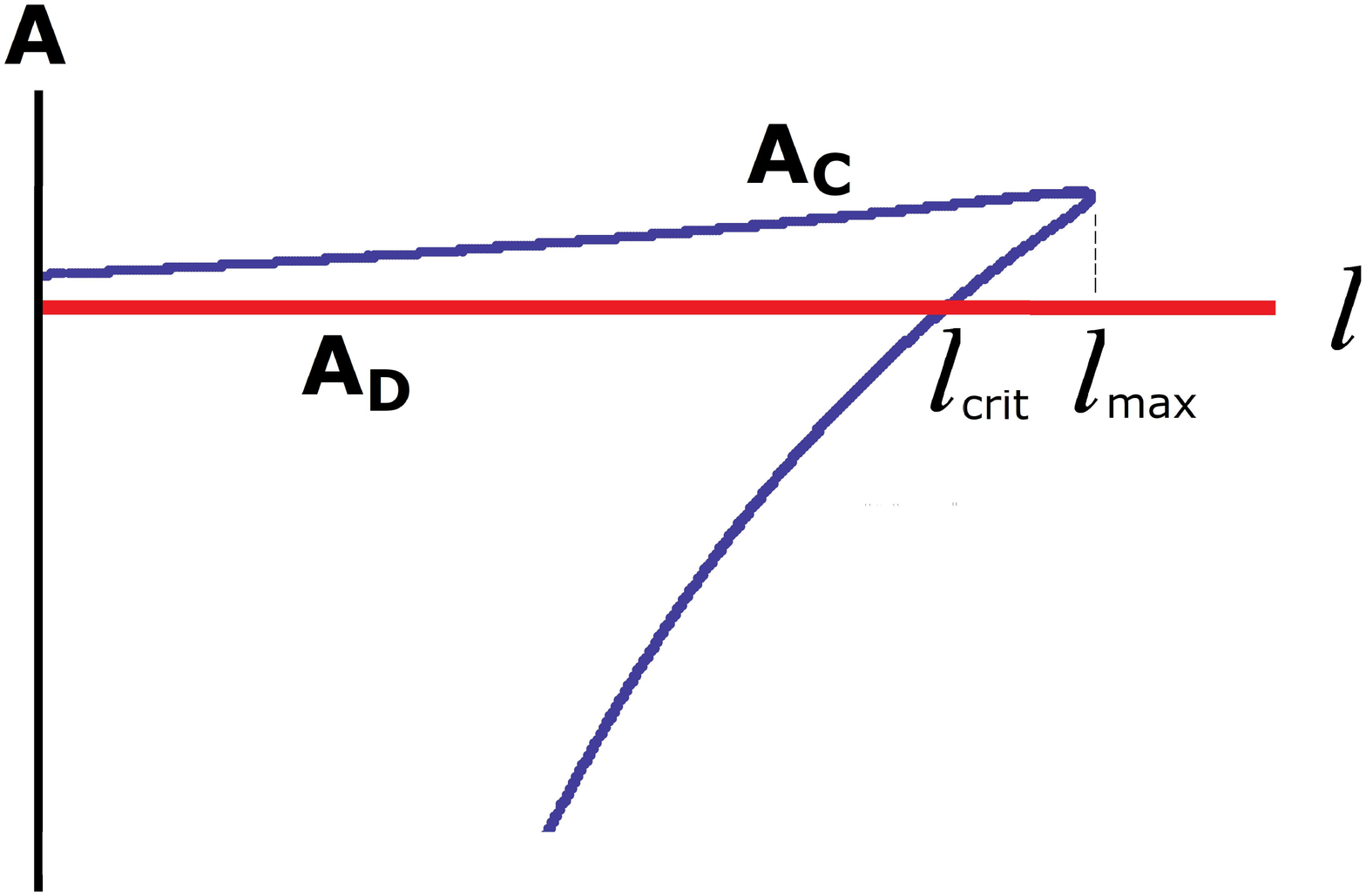}}
\\
{\it Figure 2:  The area function $A(\ell)$  :  Figure shows function $A_C(\ell)$ in the range of  $0.30\le \ell\le \ell_{max}\approx 0.47$}.  The area functions  of the connected and disconnected solutions cross at $\ell_{critical}\approx 0.45$.  $A_C(\ell_{crit})-A_D(\ell_{crit})\ne0$ and the transition is first order.}
\\

5.  To proceed let us quote following two properties: (a) As will be shown in following sections that the just replacing $L^3$ in \eqref{A} by $L^2$ we obtain the D4 on circle. (b) As that studied by Klebanov et al.  \cite{Klebanov} the phase transition is found at $\ell=1.288R_w$.  Thus, M5 system has phase transition at critical radius $R_c\equiv R_w=\ell /1.288$.  Also, using the property that $\ell$ playing the role of the inverse temperature \cite{Klebanov} we thus conclude that M5 will has phase transition at critical compactified radius.  

6.  The next problem is that if compactified M5 has phase transition then what is its low energy phase?  { First, as well-known that  the low-energy theory on the D4-branes is a perturbative 5D gauge theory.  Also, due to the Wess-Zumino coupling between the Ramond-Ramond (RR) one-form and the gauge fields on the D4-branes world-volume
\be
L_{WZ} = C_{D0}\wedge F\wedge F
\ee
a D0-brane inside a D4-brane carries the instanton charge.  In this set up the weak-coupling description of instantons is in terms of D0+D4-branes bound states. On the other hand, recent investigations had found that M5-brane on a circle admits a KK tower of massive states and these should appear as solitonic states that carry a non-vanishing instanton number \cite{Douglas,Lambert}.  Thus we have the follow two relations :
\\

${\rm Compactified~M5~with~Kaluza-Klein~modes=5D~gauge~theory~with~instanton}$\\
${\rm ~~~~~Compactified~D0+D4~at~low~energy=5D~gauge~theory~with~instanton}$
\\
\\
Therefore, the  low energy phase (i.e., large $\ell$ in figure 2) of compactified M5 shall be the compactified D0+D4.  From this point of view we thus analyze the holographic entanglement entropy of D0+D4 system in next subsection.} 
%%%%%%%%%%%%
%%%%%%%%%%%%
%%%%%%%%%%%%
\subsection {Holographic entanglement entropy of  D0+D4 on circle}
Near-horizon D0+D4 matric on a circle of coordinate $w$ with radius $R_w$ is
\be ds^2= \sqrt { R_0^3\over R_4^3}\Big(-dt^2+dx^2 +\sum_{i=1}^2 dx_i^2\Big) +{U^3\over\sqrt { R_4^3R_0^3}}\Big(1-{b^3\over U^3}\Big)dw^4+ {\sqrt{R_0^3 R_4^3}\over U^3} \Big[{dU^2\over 1-{b^3\over U^3}}+  U^2 d\Omega_4^2\Big]\nn\\
\ee
in which $b=\Big({2 \over  3R_w}\Big)^{1/2} \Big(R_0^3R_4^3\Big)^{1/4}$  and $R^3_0$, $ R^3_4$ are proportional to number of D0, D4 branes respectively. In terms of $U=4R_4^3/z^2$ and define  
\be z_c\equiv \Big({64R_4^9\over b^3}\Big)^{1/6}=2\Big({3R_w\over2}\Big)^{1/4}{R_4^{9/8}\over R_0^{3/8}}\ee
then the induced metric on a constant time slice is 
\be ds_{ind}^2=\sqrt { R_0^3\over R_4^3}\Big[\sum_{i=1}^2 dx_i^2 +\Big(1+{z'^2\over (1-{z^6\over z_c^6})}\Big)dx^2\Big]+\Big(1-{{z^6\over z_c^6}}\Big){64R_4^9\over \sqrt{R_0^3R_4^3}}{dw^2\over z^6}+ {\sqrt{R_0^3R_4^3}\over 4R_4^3}z^2d\Omega_4^2\nn
\ee
Using $\int dx_w=2\pi R_w$, $\int dx_i =L$ and $\int \Omega_4=\omega_4$ we find that surface is 
\be  A&=&\int_0^{2\pi R_w} dx_w ~\int \Pi_{i=1}^{2}dx_i~\int d\Omega_4~\int_{-\ell/2}^{\ell/2} dx~e^{-2\Phi}\sqrt {g_{ind}}\nn\\
&=&32~R_4^9~2\pi R_w~L^3~2\omega_4\int_0^{\ell/2} dx~{\sqrt{z'^2+1-{z^6\over  z_c^6}}\over z^5}
\ee
which is just that in M5 (i.e. \eqref{M5}), except that $R_5$ is replaced by $R_4$. { Thus there is a phase transition as that plotted in figure 2 while with different scale.}  Let us analyze the phases at small $\ell$  and large $\ell$ respectively in below.

First, for small $\ell$ the connected surface dominates and 
\be  A_C\approx - 32R_4^9~2\pi R_w~L^3~2\omega_4~{1\over 4 z_*^4}+~div.
\ee
As before,  because $z_c$ and $z_*$ are independent of $R_4$ and $R_4$ only appears as an overall factor $R_5^9$ before surface formula $A_C$.  Thus the holographic entanglement entropy of  compactified D0+D4 has the simple property :
\be S\sim N^3, ~~~at~small~\ell
\ee
 Since that  in this case  $R^3_4$ is proposition to number of D4, $N$.  Note that after considering the $g_s$ factor in $G_N^{(10)}$ we see that above result just produce the M5 brane entanglement entropy because the $g_s$ is the M-theory radius of  circle \cite{Gubser9602,Klebanov9604}. 

 Now,  increasing the value of $z_*$ the connected surface $A_C$ will be increasing too. And eventually, the disconnected surface will dominate and
\be A_D&=&- (32R_4^9~2\pi R_w~L^3~2\omega_4)~ {1\over 4 z_c^4}~+div.\nn\\
&=&-(32R_4^9~2\pi R_w~L^3~2\omega_4){R_0^{3/2}\over 4 2\Big({3R_w\over2}\Big){R_4^{9/2}}}+~div.
\ee
In the case of  $k$ smea D0 per unit D4 we see that at large $\ell$ 
\be S\sim N^3 {(kN)^{1/2}\over (N^{3/2})}= k^{1/2}N^2, ~~~at~large ~\ell
\ee
In conclusion, we  have investigated the holographic entanglement entropy of D0+D4 system on a circle and see that it scales as  $N^2$ at low energy, likes as a 5D gauge theory with instantons.   However, at high energy it will transform to a phase which scales as $N^3$, like as a M5 system. 
%%%%%%%%%%%%%%%%%%%%
\section {Holographic Entanglement Entropy of Compactified Dp and M Branes}  
The metric of black  Dp branes, M5 and M2 branes are
\be ds_{Dp}^2 &=& H_p^{-1/2}\Big[-F_pdt^2+dx^2+dx_0^2+\sum_{i=1}^{p-2} dx_i^2\Big]+H_p^{1/2}~\Big[ {dU^2\over F_p}+U^2~d\Omega_{8-p}^2\Big]\\
&&e^{-2\Phi}= H_p^{(p-3)/2},~~H_p=1+{R_p^{7-p}\over U^{7-p}},\\
&&F_p=1-{b^{7-p}\over U^{7-p}},~~T_H={7-p\over4\pi}b^{5-p\over 2}R_p^{p-7\over 2}\\
ds_{M5}^2 &=&H_{M5}^{-1/3}\Big[-F_{M5}dt^2+dx^2+dx_0^2+\sum_{i=1}^3 dx_i^2\Big]+H_{M5}^{2/3}~\Big[ {dU^2\over F_{M5}}+U^2~d\Omega_4^2\Big]\\
&&H_{M5}=1+{R_{M5}^3\over U^3},~~F_{M5}=1-{b^3\over U^3},~~T_H={3\over4\pi}b^{1\over 2}R_{M5}^{3\over 2}\\
ds_{M2}^2 &=&H_{M2}^{-2/3}\Big[-F_{M2}dt^2+dx^2+dx_0^2\Big]+H_{M2}^{1/3}~\Big[ {dU^2\over F_{M2}}+U^2~d\Omega_7^2\Big]\\
&&H_{M2}=1+{R_{M2}^6\over U^6},~~F_{M2}=1-{b^6\over U^6},~~T_H={3\over2\pi}b^{2\over 2}R_{M2}^{3}\ee
in which the Hawking temperature $T_H$ is determined at near-horizon limit.   Using above relations we can obtain metric of  associated branes compactified on coordinate $w$ with radius $R_w= 1/2\pi T_H$ 
\be ds_{Dp}^2 &=& H_p^{-1/2}\Big[-dt^2+F_pdw^2+dx^2+\sum_{i=1}^{p-2} dx_i^2\Big]+H_p^{1/2}~\Big[ {dU^2\over F_p}+U^2~d\Omega_{8-p}^2\Big]\\
ds_{M5}^2 &=&H_{M5}^{-1/3}\Big[-dt^2+F_{M5}dw^2+dx^2+\sum_{i=1}^3 dx_i^2\Big]+H_{M5}^{2/3}~\Big[ {dU^2\over F_{M5}}+U^2~d\Omega_4^2\Big]\\
ds_{M2}^2 &=&H_{M2}^{-2/3}\Big[-dt^2+F_{M2}dw^2+dx^2\Big]+H_{M2}^{1/3}~\Big[ {dU^2\over F_{M2} }+U^2~d\Omega_7^2\Big]
\ee
To consider the entanglement entropy of the ``slab" in which the subspace is defined by $-\ell /2 < x < \ell/2$  we need to investigate  the following induced metric
\be
ds_{Dp,ind}^2&=&H_p^{-1/2}\Big[F_pdw^2+\sum_{i=1}^{p-2} dx_i^2\Big]+(H_p^{-1/2}+H_p^{1/2}{U'^2\over F_p})dx^2+U^2~H_p^{1/2}d\Omega_{8-p}^2\nn\\
\\
ds_{M5,ind}^2 &=&H_{M5}^{-1/3}\Big[F_{M5}dw^2+\sum_{i=1}^3 dx_i^2\Big]+(H_{M5}^{-1/3}+ H_{M5}^{2/3}{U'^2\over F_{M5}})dx^2+U^2~H_{M5}^{2/3}~d\Omega_4^2\nn \\
\\
ds_{M2,ind}^2 &=&H_{M2}^{-2/3}\Big[F_{M2}dw^2+dx^2\Big]+(H_{M2}^{-2/3}+H_{M2}^{1/3} {U'^2\over F_{M2} })dx^2+U^2~H_{M2}^{1/3}d\Omega_7^2\Big]
\ee

Thus, the associated entanglement entropy is
\be S_{A}^{(Dp)}&=&{1\over 4G_N^{(10)}}\int_\gamma d^8\sigma~\sqrt{H_p~F_p}~ U^{8-p}\sqrt{1+H_p{U'^2\over F_p}} \\
S_A^{(M5)}&=&{1\over 4G_N^{(11)}}\int_\gamma d^8\sigma~\sqrt{H_{M5}~F_{M5}}~U^{4}\sqrt{1+H_{M5}{U'^2\over F_{M5}}}\\
S_A^{(M2)}&=&{1\over 4G_N^{(11)}}\int_\gamma d^8\sigma~\sqrt{H_{M2}~F_{M2}}~U^{7}\sqrt{1+H_{M2}{U'^2\over F_{M2}}}
\ee
From the definitions of $H_p$, $H_{M5}$, $H_{M2}$, $F_p$, $F_{M5}$ and $F_{M2}$ we see the property that the area of surfaces of M5 and D4 branes have the similar  function form and that of M2 and D1 branes have the similar  function form too.  Thus, M5 has the same phase structure as D4 and  M2 has the same phase structure as D1.  Note that in the case of $F_{p}=F_{M5}=F_{M2}=1$, which is the zero-temperature system without compactification, above property also useful.  Also, remaining $F_A$ in ${U'^2\over F_A}$ while let $F_A=1$ in $H_A~F_A$, which is the finite-temperature system without compactification, above property also useful.  Thus, in following section we merely analyze the property of $S_{A}^{(Dp)}$.

In a similar way, the metric of Dp branes smea on black $D_{p+4}$  is 
\be  ds^2&=&H_p^{-1/2}H_{p+4}^{-1/2}\Big[-F dt^2+\sum_1^2 dy_i^2\Big]+H_p^{1/2}H_{p+4}^{-1/2}\Big[dx^2+dx_0^2+\sum_1^2 dx_i^2\Big]\nn\\
&&+H_p^{1/2}H_{p+4}^{1/2}~\Big[ {dU^2\over F_p}+U^2~d\Omega_{4-p}^2\Big]\\
&&H_{p,p+4}=1+{R_{p,p+4}^{3-p}\over U^{3-p}},~~F=1-{b^{3-p}\over U^{3-p}},\\
&&T_H={3-p\over4\pi}{b^{2-p}\over (R_{p}R_{p+4})^{3-p\over2}},~~e^{-2\Phi}=H_p^{(p-3)/2}H_{p+4}^{(p+1)/2}
\ee
in which the Hawking temperature $T_H$ is determined at near-horizon limit.   Using above relations we can obtain metric of  associated branes compactified on coordinate $w$ with radius $R_w= 1/2\pi T_H$
\be
ds^2&=&-H_p^{1/2}H_{p+4}^{-1/2} dt^2+H_p^{-1/2}H_{p+4}^{-1/2}F dw^2+H_p^{1/2}H_{p+4}^{-1/2}\Big(dx^2+\sum_1^2 dx_i^2\Big)\nn\\
&&+H_p^{1/2}H_{p+4}^{1/2} \Big({dU^2\over F}+U^2 d\Omega_{4-p}^2\Big)
\ee
 The induced metric in considering the entanglement entropy of the ``slab" in which the subspace is defined by $-\ell /2 < x < \ell/2$ is
\be
ds_{ind}^2&=&H_0^{1/2}H_4^{-1/2}\Big(\sum_1^2 dx_i^2\Big)+H_0^{-1/2}H_4^{-1/2}F dw^2+\Big(H_0^{1/2}H_4^{-1/2}+H_0^{1/2}H_4^{1/2}{U'^2\over F}\Big) dx^2\nn\\
&&+H_0^{1/2}H_4^{1/2} U^2 d\Omega_4^2
\ee
Thus, the associated entanglement entropy is
\be S_{A}^{(Dp+D_{p+4})}&=&{1\over 4G_N^{(10)}}\int_\gamma d^8\sigma~\sqrt{H_p~F_p}~ U^{8-p}\sqrt{1+H_p{U'^2\over F_p}} 
\ee
Which is just the function of $S_{A}^{(Dp)}$ as claimed before in the case of D0+D4. However we shall notice that the Hawking temperature $T_H$ and the associated radius of compactified coordinate $R_w$ between ${\rm D_p}$ brane system and ${\rm D_p+D_{p+4}}$ brane system are different from each other.

%%%%%%%%%%%%%%%%%%
\section {Phase Transition of Compactified Branes}
\subsection{General formulation}  
From above section we see that in near-horizon lime the entanglement entropy of Dp brane can be expressed as
\be S_{A}^{(Dp)}&\approx&{1\over 4G_N^{(10)}}\int d^8\sigma~\sqrt {R^{7-p}\over U^{7-p}}~ U^{8-p}~\sqrt{1-{b^{7-p}\over U^{7-p}}+{R^{7-p}\over U^{7-p}}~U'^2}\nn\\
&=&{1\over 4G_N^{(10)}}R^{(7-p)^2\over 5-p}~\int d^8\sigma~\Big({|5-p|\over 2}~z\Big)^{9-p\over p-5}~\sqrt{1-{z^{2(7-p)\over 5-p}\over  z_c^{{2(7-p)\over 5-p}}}+z'^2}
\ee
in which we define
\be U^{-{5-p\over 2}}&\equiv& {|5-p|\over 2}R^{-{7-p\over 2}}~z\\
z_c^{{2(7-p)\over 5-p}}&\equiv&{\Big({|5-p|\over 2}\Big)^{2(7-p)\over p-5}~R^{(7-p)^2\over 5-p}\over b^{7-p}}=\Big({7-p\over |5-p|}~R_w\Big)^{2(7-p)\over 5-p}
\ee

We now need to find the minimal surface by minimizing the above area functional.  Let $z'=0$ at the  turning point of the minimal surface, i.e. $z=z_*$, then
\be z'&=& \sqrt{{z_*^{2\beta}\over {z}^{2\beta}}{\Big(1-\Big({z\over z_c}\Big)^\alpha \Big)^2\over 1-\Big({z_*\over z_c}\Big)^\alpha }-\Big(1-\Big({z\over z_c}\Big)^\alpha\Big)}
\ee
Thus
\be
\label{4.5}
\ell &=& 2 \int_0^{\ell/2}dx =2 \int_0^{z_*}dz {dx\over dz}\nn\\
&=&2 \int_0^{z_*}dz {z^\beta\sqrt{1-\Big({z_*\over z_c}\Big)^\alpha}\over \sqrt{z_*^{2\beta}\Big(1-\Big({z\over z_c}\Big)^\alpha \Big)^2-{z}^{2\beta}\Big(1-\Big({z\over z_c}\Big)^\alpha\Big)\Big(1-\Big({z_*\over z_c}\Big)^\alpha\Big)}}
\ee
and
\be
\label{4.6}
A&=& {1\over 4G_N^{(10)}} \Big({|5-p|\over 2}\Big)^{9-p\over p-5} ~2\pi R_w~L^{p-2}~\omega_{8-p}\int_{-\ell/2}^{\ell/2} dx ~{1\over z^\beta} \sqrt {1-\Big({z\over z_c}\Big)^\alpha +z'^2}\nn\\
&=&{1\over 4G_N^{(10)}}\Big({|5-p|\over 2}\Big)^{9-p\over p-5} ~4\pi R_w~L^{p-2}~\omega_{8-p}\int_0^{z_*} dz ~{z_*^\beta\over z^\beta}{{\sqrt{1-\Big({z\over z_c}\Big)^\alpha}\over \sqrt {z_*^{2\beta }\Big(1-\Big({z\over z_c}\Big)^\alpha \Big)- z^{2\beta}\Big(1-\Big({z_*\over z_c}\Big)^\alpha \Big)}}}\nn\\
\ee
in which 
\be \alpha={2(7-p)\over 5-p},~~~\beta ={9-p\over 5-p},~~~
\int \sum_{i=1}^{p-2} dx_i=L^{p-2},~~~\int d\Omega_{8-p}=\omega_{8-p}
\ee
 In numerical analysis, we can use \eqref{4.5} to find the function $\ell (z_*)$ and use \eqref{4.6} to find the surface function $A(z_*)$ as before.  Then, we use the two functions to obtain the surface function $A(\ell)$.  However, we will use the analytic method to prove that there always is phase transition in these systems.
\subsection{Analysis }  
First, from (4.6) we see that  in the cases of $z_* \approx 0$ the connected surface is 
\be A_{con} &\approx&{1\over 4G_N^{(10)}}\Big({|5-p|\over 2}\Big)^{9-p\over p-5} ~4\pi R_w~L^{p-2}~\omega_{8-p}\int_\epsilon^{z_*} dz ~{1\over z^\beta}\nn\\
&\approx&{1\over 4G_N^{(10)}}\Big({|5-p|\over 2}\Big)^{9-p\over p-5} ~4\pi R_w~L^{p-2}~\omega_{8-p}~\Big[{1\over -\beta+1}\Big(z_*^{-\beta+1}-\epsilon^{-\beta+1}\Big) \Big]
\ee
As the disconnected surface is defined by
\be A_{dis} &\equiv&{1\over 4G_N^{(10)}}\Big({|5-p|\over 2}\Big)^{9-p\over p-5} ~4\pi R_w~L^{p-2}~\omega_{8-p}\int_\epsilon^{z_c} dz ~{1\over z^\beta}\nn\\
&=&{1\over 4G_N^{(10)}}\Big({|5-p|\over 2}\Big)^{9-p\over p-5} ~4\pi R_w~L^{p-2}~\omega_{8-p}~\Big[{1\over -\beta+1}\Big(z_c^{-\beta+1}-\epsilon^{-\beta+1}\Big) \Big]
\ee
thus
\be A_{con} -A_{dis}&=&{1\over 4G_N^{(10)}}\Big({|5-p|\over 2}\Big)^{9-p\over p-5} ~4\pi R_w~L^{p-2}~\omega_{8-p}~\Big[{1\over -\beta+1}\Big(z_*^{-\beta+1}-z_c^{-\beta+1}\Big) \Big]\nn\\
&=&{1\over 4G_N^{(10)}}\Big({|5-p|\over 2}\Big)^{9-p\over p-5} ~4\pi R_w~L^{p-2}~\omega_{8-p}~z_*^{-\beta+1}~\Big[{1-{(z_c/z_*)^{-\beta+1}}\over -\beta+1}\Big]
\ee
Note that $z_c >z_*$ and $-\beta+1={4\over p-5}$ which may be positive or negative. However, in any case the value $A_{con} -A_{dis}$ is always negative and we conclude that the connected surface dominates at small vale of $z_*$.

Now, when we increasing $z_*$ to $z_c$ the connected surface will be  increasing to the area of $A_{dis}$.  If at $z_*=z_c$ the slop of  $A_{con}$ is a decreasing function then we can conclude that $A_{con} -A_{dis} >0$ in some region of $z_*\approx z_c$. Thus, the  disconnected surface will dominate in this region and we have a phase transition. This is the case we will prove. 

Through the calculation we find that  
\be
 {dA_{con}\over dz_*}&=&{1\over 4G_N^{(10)}}\Big({|5-p|\over 2}\Big)^{9-p\over p-5} ~4\pi R_w~L^{p-2}~\omega_{8-p}~\Big[{1\over z_*^\beta}-{\alpha\over 2z_c^{2\beta+1}}\int_0^{z_*}dz~{z^\beta\over 1-z^\alpha z_c^{-\alpha}}\Big]\nn\\
  &=&{1\over 4G_N^{(10)}}\Big({|5-p|\over 2}\Big)^{9-p\over p-5} ~4\pi R_w~L^{p-2}~\omega_{8-p}~\Big[{1\over z_*^\beta}+{1\over 2z_c^\beta}~\ell n\Big(1-{z_*\over z_c}\Big)\Big]
\ee
which implies that
\be
\Rightarrow&&{dA_{con}\over dz_*}{\mid_{z_*=z_c}}=-\infty,
\ee
Thus, near the region $z_*\approx z_c$ the area of connected surface is larger than that of the connected surface and the system will transform to disconnected surface.
%%%%%%%%%%%%%%%%%%
\section {Conclusion}  
In this paper we use the entanglement entropy to study the phase transition of M-branes and $D_p$ branes systems. We first  show that the entanglement entropy of  M5 branes on a circle  scales as $N^3$ \cite{Gubser9602,Klebanov9604} and has the phase transition during decreasing the compactified radius.  We next show that  the holographic entanglement entropy of D0+D4 system on a circle will scale as  $N^2$ at low energy, likes as a  gauge theory with instantons.  However, at high energy it  transforms to a phase which scales as $N^3$, like as the  M5 branes system. The property is consistent with  \cite{Douglas, Lambert} which showed that the compactified M5 with Kaluza-Klein modes is just the 5D gauge theory with instanton. Thus the  low energy phase of compactified M5  is the compactified D0+D4 which becomes 5D gauge theory with instanton at low energy. We have also seen that the entangle entropy of compactified  M5, M2  and ${\rm D_p+D_{p+4}}$  has similar mathematical form likes as that of  Dp branes. Use this property we then prove that they all have phase transition from connected to disconnected surface during increasing the line segment of length $\ell$ which dividing the space. 

{ Finally, we  make following comments to conclude this paper. 

 1. The analysis in this paper are in the leading order of $N$ approximation. It is interesting to investigate the problem beyond the leading order approximation to see whether the phase will still happen at finite $N$ and, could the the transition become second order at small $N$ ? 

 2. It is interesting to use  Hartnoll et al. method \cite{Hartnoll} to study the  phase transition in the compactified branes and compare it with this paper. 

 3. As that mentioned in the introduction the Renyi entropy is a one-parameter generalization of entanglement entropy labeled by an integer $n$.  Thus it is interesting to study the problem of how the phase transition in the compactified branes will depend on $n$. 

 These questions are reserved for future research.}
\\
\\
%%%%%%%%%%%%%%%%%%%%%%
\begin{center} {\bf REFERENCES}\end{center}
%%%%%%%%%%%%%%%%%%%%%%

\end{document}